Title

Statistical Methodologies for Urban Morphology Indicators: A Comprehensive Review of Quantitative Approaches to Sustainable Urban Form

**Author:** Mahshid Gorjian

**Affiliation:** University of Colorado Denver

**Emai:** Mahshid.gorjian@ucdenver.edu

**ORCID:** https://orcid.org/0009-0000-9135-0687

**Correspondent Author:** Mahshid Gorjian

**Abstract**

Urban morphology has a significant impact on sustainable urban outcomes in the ecological, social, and economic sectors. Nonetheless, academics lack clarity and consensus on how to quantify urban morphology indicators, which impedes multidisciplinary research and the creation of integrated sustainable solutions. This comprehensive study develops a detailed taxonomy of urban form indicators related to sustainability, describes quantitative approaches for investigating the linkages between urban morphology and sustainability, and examines research trends, gaps, and future directions. A review of 89 carefully selected studies showed 365 distinct urban morphology measures, with significant variation in names, interpretations, and data formats. The measures were divided into six categories: urban tissue configuration, roadway network, building plot characteristics, land use, natural elements and greenspace, and urban expansion, each with its own subclass. The study indicates that, even though urban morphology is intrinsically spatial, spatial modeling is rarely used to investigate its relationships with sustainability. The research shows a regional bias, with insufficient representation of studies focusing on the Global South and natural elements. This synthesis calls for a standardized classification framework as well as more broad, spatially explicit techniques to foster multidisciplinary research and support effective policy and planning for sustainable urban development.

**Keywords**

Sustainability; Urban Morphology; Landscape; Measures; Spatial Analysis; Quantitative Methods

**Introduction**

With more than half of the world's population now living in cities, urbanization continues to affect landscapes around the world. Rapid urbanization has complex and interwoven effects on the environment, economy, and society (Hu et al., 2021; Purvis et al., 2019). As cities grow, it becomes more challenging to balance resource usage, environmental sustainability, social unity, and economic prosperity. The difficulties make the pursuit of sustainability, in which humans and nature coexist peacefully for current and future generations, particularly vital in the urban context.

Cities contain dynamic systems that contribute to ecosystem degradation while also encouraging socioeconomic development, technological innovation, and cultural shift (Wu, 2014). Effective solutions for sustainable urban development require a thorough grasp of the dynamics and transitions of urban systems. Planning and policy decisions require cross-disciplinary ideas as well as reliable, evidence-based information (Batty, 2012). Urban morphology, the physical form of cities, and its impact on long-term outcomes are an exciting but underexplored subject (Albert, 1999; Jabareen, 2006).

Urban morphology examines the layout, organization, and spatial distribution of urban elements such as structures, parcels, thoroughfares, and public spaces, as well as the dynamics of human interaction with these environments across time (Kropf, 2018; Moudon, 1997). The spatial arrangement of urban structures can have an impact on environmental conditions, transportation options, energy consumption, social interactions, and economic prospects. Recent studies have found links between urban morphology and specific sustainability outcomes, such as air quality (Kang et al., 2019), energy consumption (Rode et al., 2014), greenhouse gas emissions (Liu et al., 2017), transportation behaviors (Rybarczyk & Wu, 2014), urban vitality (Tang et al., 2018), and land consumption and growth patterns (Cheng, 2011; Milad et al., 2018). These studies usually use quantitative and spatially explicit methods to investigate the relationships between urban form and sustainability. Nonetheless, the results are incongruent due to differences in context, geographical emphasis, and assessment methodology.

A major impediment to progress is a lack of clarity and consensus on the definition and assessment of urban morphology. The literature shows significant differences in language, operational definitions, and indicator selection (Fleischmann et al., 2021; Jabareen 2006). The term "compactness," a frequently studied attribute, has been defined in a variety of ways, including population density, spatial clustering, and building scale, leading to conflicting conclusions about its impact on travel behavior, social cohesion, and environmental performance (Nam et al., 2012; Hemani et al., 2017; Kang et al., 2019). Other commonly used indicators exhibit comparable variability, complicating comparative research and reducing the reliability of policy recommendations.

This mismatch reflects the diverse nature of urban sustainability research, which draws on concepts and approaches from disciplines such as ecology, geography, public health, transportation engineering, and landscape architecture. Despite advances in geographic information systems (GIS), remote sensing, and data science that enable complex spatial studies at several scales (Batty & Longley, 1994; Taubenböck et al., 2020), researchers are frequently constrained by disciplinary boundaries. The absence of established measurements and a unified methodology reduces the reproducibility of outcomes and impedes interdisciplinary collaboration, ultimately compromising the evidentiary foundation for sustainable urban planning and administration.

Correcting these inadequacies is critical for several reasons. Cities worldwide are under increasing strain from rapid urbanization, climate change, and environmental degradation, needing efficient and scalable responses. Second, planners and politicians want consistent, transparent frameworks for evaluating urban forms in different situations and developing sustainable development plans. Third, as urban sustainability becomes a primary priority in global research and policy frameworks, developing interdisciplinary consensus will help promote common goals.

This analysis addresses the challenges by focusing on sustainable urban morphology, a field that studies the relationship between cities' spatial patterns and structures and their environmental, social, and economic sustainability. The review seeks to give a complete synthesis of the measurements, methodology, and research trends in this topic, with the following key objectives:

1. Develop a comprehensive taxonomy of quantitative measures for urban morphology that relate to environmental, economic, and social sustainability outcomes.

2. Carefully describe the quantitative and geographical analysis tools used to investigate the relationships between urban morphology and sustainability.

3. To identify current research trends, highlight knowledge gaps, and suggest future routes for quantifying urban morphological characteristics to promote sustainable urban development.

**Methods**

This review followed a methodical approach to ensure transparency, replicability, and comprehensive coverage of relevant literature. The search strategy centered on six major academic databases: Sage, Scopus, ScienceDirect, Wiley Online Library, Taylor & Francis, and ProQuest, which were chosen for their extensive indexing of papers in urban planning, environmental science, landscape design, and urban geography. The search included Boolean operators and wildcard terms, using the syntax: "urban morphology" OR "urban form" OR "urban pattern" OR "urban fabric" AND "spatial*" AND "sustainab*". The time scope included all available years up to the review date, and searches were conducted on articles published in English. Conference proceedings, book chapters, and reports were excluded to focus on peer-reviewed, full-text journal publications.

The inclusion criteria required that studies (1) use geographical data and analyses related to urban morphology, (2) use quantitative or mixed-method techniques at the building scale or higher, and (3) explicitly link urban morphology metrics to at least one sustainability outcome. Exclusion criteria excluded studies that used only qualitative methodologies (e.g., content analysis or historical interpretation), those that lacked quantitative modeling of urban form-sustainability relationships, and publications that were not completely accessible or published in English.

The selection technique began with database searches, which yielded 447 preliminary records. After eliminating duplicates, 350 articles were retained for screening. The use of inclusion and exclusion criteria during title, abstract, and keyword screening reduced the pool to 132 articles suitable for full-text examination. Following the extensive assessment, 89 publications were selected for final analysis based on their relevance, methodological integrity, and adherence to inclusion criteria. A PRISMA flow diagram depicted the selection process and ensured transparency on the number of records identified, examined, included, and deleted at each stage (Liberati et al. 2009).

The data extraction technique included categorizing each article by publication year, journal, research region, spatial data type, unit of analysis, analytical methodologies, urban morphology element, particular indicators and their definitions, and the sustainability outcomes studied. Descriptive data revealed research tendencies based on year, geographic emphasis, and journal domain. The use of content analysis and a meta-synthesis strategy facilitated the discovery and consolidation of themes, metrics, and analytical techniques. Kropf (2018) defined urban morphology as six basic aspects: urban tissue configuration, roadway network, building-plot features, land use, natural elements and greenspace, and urban expansion. The review used this theoretical framework. Each facet included applicable subcategories, allowing for a thorough taxonomy of the 365 unique metrics identified in the literature.

The synthesis examined the distribution of different metrics and methodologies across areas, sustainability topics, and time periods. The review underlined the widespread use and execution of critical metrics, the integration of spatial and aspatial models, and the diversity of data sources.

Numerous constraints influenced the search and selection process. Limiting the search to English-language literature likely excluded relevant research published in other languages, particularly those from the Global South. The reliance on peer-reviewed publications overlooked grey literature and alternative sources that could provide valuable information. The sole emphasis on quantitative and mixed-method studies, while improving rigor and comparability, prohibited qualitative research that could have provided important background or theoretical insights. Despite these limits, the study covered a wide range of current research on sustainable urban morphology.

**Thematic/Topical Sections**

The structure of urban tissue, including the spatial organization and characteristics of created settings, has a significant impact on urban sustainability outcomes. Factors like as spatial distribution, density, shape, and patch properties frequently lead analysis in this subject. Recent research has demonstrated that parameters such as population density, built-up area density, and spatial clustering have a significant impact on air quality, urban expansion, and land consumption trends (Kang et al., 2019; Cheng, 2011).

Research generally agrees that high-density, compact urban designs reduce land use and encourage more efficient infrastructure utilization. Compactness relates to shorter commute times and increased transportation efficiency (Song et al., 2017; Ye et al., 2015). Nonetheless, certain investigations yield inconsistent results. Spatial statistics, such as Moran's I, show that compactness correlates favorably with increased vehicular trips and ozone levels, highlighting the complexities of these relationships. Population density can increase social cohesion (Hemani et al., 2017) while decreasing active leisure time (Fan, 2010), highlighting the relevance of scale and context.

Inconsistency in operational definitions is a significant shortcoming in this body of work. Metrics commonly differ in nomenclature and computation, making cross-study comparisons and generalizability difficult (Fleischmann et al., 2021). Despite advances in remote sensing and GIS technologies that enable more precise mapping, the subject still lacks defined procedures for spatial delineation and measurement.

Research highlights the need of multi-scalar and multi-faceted methodologies, particularly those that use spatial and temporal data, in fully understanding the dynamics of urban fabric and its sustainability implications.

The design and quality of roadway networks have an impact on mobility and accessibility, and as a result, sustainability outcomes such as transportation emissions and urban vibrancy. This domain's key metrics include road density, network connectivity, space syntactic parameters, and street design features.

There is widespread consensus that well-connected, intricate roadway networks promote sustainable transportation options such as walking, cycling, and public transportation (Boeing, 2018; Rybarczyk & Wu, 2014). These networks promote social justice and accessibility, as evidenced by mixed-method research from India and Europe, which found that connectivity and greenspace accessibility were associated with increased social cohesion and well-being (Hemani et al., 2017; Sánchez et al., 2019).

Nonetheless, the study points to major disparities in the assessment and implementation of connectivity and density. Calculations of words such as "road network density" may differ depending on the spatial unit or input data used (Yuan et al., 2018; Rybarczyk and Wu, 2014). Despite providing a theoretically sound framework for measuring accessibility and visibility in street networks, space syntax is still limited to certain regions and fields.

A notable gap in this research topic is the underuse of three-dimensional network measurements that account for building height and street canyon effects, which could improve knowledge of urban microclimate and energy consumption. The emphasis on two-dimensional metrics hampers understanding of the overall consequences of urban connectivity.

In conclusion, street network research consistently acknowledges connectivity as a critical aspect for long-term results; yet methodological differences and the limited use of advanced measures prevent more robust comparison analysis.

The physical characteristics of buildings and land, such as density, height, direction, usage, and parcel attributes, have a significant impact on urban energy consumption, microclimate, and overall sustainability. Quantitative studies have developed various measurements in these domains; nonetheless, standardization remains elusive.

Studies consistently show that increased building density and improved architectural design result in higher energy efficiency and lower heat island effects (Chatzipoulka & Nikolopoulou, 2018; Rode et al., 2014). The relationship between building features and sustainability outcomes is influenced by factors such as local climate, urban context, and land use diversity. The same density metric may yield different results depending on whether it is calculated using the number of buildings per area, building volume, or parcel size (Chatzipoulka & Nikolopoulou, 2018; Salvati, 2013).

The importance of building orientation and form continues to be debated, particularly in terms of sun access, ventilation, and microclimate control. Several studies emphasize the need for more specific, three-dimensional data, such as building volume or sky view factor; yet most of the research remains based on easily accessible two-dimensional metrics.

The use of high-resolution remote sensing and GIS data in modern research is a considerable advantage, allowing for full mapping of architectural features on a broad scale. Nonetheless, differences in data availability and quality between places limit the generalizability of findings, especially in the Global South.

To advance the subject, established definitions must be implemented, as well as sophisticated spatial datasets used more widely to permit cross-contextual comparisons.

The literature on sustainable urban morphology emphasizes land use diversity, density, and proximity. Metrics such as the Shannon entropy index quantify the diversity of residential, commercial, and industrial functions in urban areas, influencing transportation, emissions, and social equality effects.

Empirical research repeatedly shows that higher land use diversity promotes more sustainable mobility behaviors, such as increased walking and reduced reliance on private vehicles (Dur et al., 2014; Ding, 2014). Mixed-use settings are linked to enhanced urban vitality and economic benefits.

Nonetheless, errors in the computation and application of diversity metrics, particularly the Shannon index, are common. Normalization, spatial unit, and data source all have different operational definitions (Han et al., 2019). The relationship between land use mix and specific sustainability outcomes may change between cultural and socioeconomic contexts, as evidenced by research findings in China, North America, and Europe.

The literature suggests that more extensive evaluations linking land use variety with other aspects of urban planning, such as connection and accessibility, are required to improve knowledge of their cumulative impact on sustainability.

Natural features and green spaces provide vital ecosystem services in metropolitan areas, promoting biodiversity, climate regulation, and human well-being. This issue is significantly underrepresented in modern research, with only a few studies addressing the spatial arrangement, accessibility, and usefulness of urban natural features.

According to the studies evaluated, greenspace coverage, density, park accessibility, and landscape configuration all have positive connections with ecosystem health and socioeconomic sustainability (Salvati et al., 2014; Sánchez et al., 2019). However, the operationalization of these variables is highly diverse, and few studies systematically incorporate ecosystem service valuation or spatial design into sustainability assessments.

A fundamental constraint is the academic divide between urban morphology and landscape ecology, as many ecosystem-oriented studies fail to explicitly integrate urban form, and vice versa. Bridging this gap requires improved methodological and conceptual integration, as well as broader adoption of standardized landscape metrics.

As climate change and urbanization place additional strain on urban ecosystems, it is critical to increase research attention on natural elements and their contribution to sustainability.

Urban expansion patterns and border adjustments have a substantial impact on sustainability, influencing land usage, infrastructure requirements, and ecological fragmentation. Quantitative studies use metrics like urban boundary complexity, sprawl indices, and expansion rates to assess these dynamics.

The consensus is that compact growth patterns and clearly defined urban bounds support more sustainable development outcomes, such as lower resource consumption and higher service efficiency (Dong et al., 2019; Salvati, 2013). Nonetheless, methodologies for quantifying urban expansion are strikingly variable, with the majority used only in isolated research, limiting the ability to combine data from multiple contexts.

Furthermore, the research frequently favors large metropolitan areas, leaving out smaller cities and peri-urban regions where growth trajectories may diverge. The absence of longitudinal, multi-scalar analysis restricts understanding of the evolution of urban growth dynamics across time.

Improving the assessment and surveillance of urban expansion, considering both spatial and temporal dimensions, will be critical for directing future sustainable urban development.

Urban morphology research increasingly recognizes the importance of standardized, multidimensional metrics as well as increased methodological rigor. The subject is hampered by inconsistent measurement, poor integration of sophisticated spatial modeling, and insufficient emphasis on natural factors and the Global South. Numerous studies focus on the Global North, whereas the rapid urbanization occurring in the Global South receives scant attention.

**Discussion**

The comprehensive review of spatial measures and approaches in sustainable urban morphology reveals a rapidly evolving but mainly disconnected sector (Fleischmann et al., 2021; Zhang et al., 2023). The synthesis of studies across topic areas highlights the growing recognition that urban morphology, which encompasses urban tissue structure, street networks, building-plot features, land use, natural components, and urban expansion, has a direct impact on the sustainability of cities (Jabareen, 2006; Kropf, 2018; Cheng, 2011; Zhang et al., 2023). Nonetheless, considerable differences and shortcomings in measurement, technique, and study emphasis continue to impede study comparability and the practical application of research findings (Fleischmann et al., 2021; Batty, 2012).

The review's central focus is the variety and growth of urban morphology measures. Researchers discovered over 365 unique measures across six primary dimensions; however, just a few, such as landscape metrics, built-up area density, the Shannon entropy index, and space syntax indicators, have seen widespread application (Zhang et al., 2023). Even among operational definitions and computation approaches, there is significant heterogeneity (Fleischmann et al., 2021; Nam et al., 2012). The lack of uniformity makes it difficult to synthesize findings, compare results from different studies, and formulate generalizable assumptions concerning urban form and sustainability (Jabareen, 2006; Fleischmann et al., 2021; Zhang et al., 2023). This complicates the translation of research findings into policy recommendations and planning guidelines, especially for practitioners looking for clear, evidence-based methods (Batty, 2012; Jabareen, 2006).

Simultaneously, research in various theme areas shows that some urban design patterns consistently support sustainable outcomes. High-density, compact urban design reduce land use and promote sustainable transportation, whereas interconnected street networks provide access to opportunities and reduce dependency on private autos (Song et al., 2017; Boeing, 2018; Ye et al., 2015; Dur et al., 2014; Rybarczyk & Wu, 2014). Building plot characteristics, such as density and orientation, influence energy consumption and urban microclimates; nonetheless, varied and cohesive land use patterns encourage urban vibrancy and economic sustainability (Chatzipoulka & Nikolopoulou, 2018; Rode et al., 2014; Salvati, 2013; Ding, 2014). The favorable effects of greenspace and natural features on ecosystem health and societal well-being, albeit less frequently studied, are widely documented in the literature (Salvati et al., 2014; Sánchez et al., 2019).

Despite these advances, the review highlights some substantial limitations within the current body of research. Numerous studies rely on exploratory geographical analysis and aspatial statistical approaches, whereas spatial modeling, which can address spatial autocorrelation and heterogeneity, is overlooked (Batty & Longley, 1994; Zhang et al., 2023). This limits the robustness and policy implications of empirical findings. Furthermore, the prevalence of two-dimensional metrics, particularly in the analysis of street networks and buildings, restricts researchers' ability to fully comprehend the complexities of urban form, particularly three-dimensional elements that influence microclimate, energy consumption, and human experience (Chatzipoulka & Nikolopoulou, 2018; Batty & Longley, 1994).

The field is defined by fragmentation and disciplinary isolation. Research in urban morphology typically fails to integrate with advances in landscape ecology, environmental psychology, and urban economics, impeding the development of truly integrated models of sustainable urban design (Batty, 2012; Fleischmann et al., 2021). Ecosystem services provided by natural features in urban environments are rarely included in traditional urban morphology studies, despite their growing importance in policy and research (Salvati et al., 2014; Zhang et al., 2023). The continued use of imprecise terminology for compactness, connectivity, and land use mix exacerbates conceptual confusion and impedes interdisciplinary collaboration (Fleischmann et al., 2021; Nam et al., 2012).

The literature is primarily concentrated on the Global North, with 85% of studies focused on Europe, North America, and China (Zhang et al., 2023). The most rapidly urbanizing regions, such as Africa, South Asia, and Latin America, are significantly underrepresented (Hu et al., 2021; Zhang et al., 2023). Language barriers and limited access to high-resolution spatial data increase this discrepancy, as does the removal of non-English and non-peer-reviewed articles from numerous systematic studies (Zhang et al., 2023). As a result, the urban morphology measurements and sustainable results reported in the literature may not sufficiently reflect the unique problems and opportunities faced by rapidly expanding cities in the Global South (Hu et al., 2021; Zhang et al., 2023).

Significant information gaps remain, notably in the integrating of natural aspects and ecosystem vitality into urban morphology research (Salvati et al., 2014; Zhang et al., 2023). Although data shows that greenspace has a positive impact on biodiversity, climate adaptation, and human well-being, few studies study these linkages with the same rigor as those on the built environment (Salvati et al., 2014; Sánchez et al., 2019). Assessments of ecosystem services, landscape connectivity, and environmental quality are inconsistently included, and when they are, the conceptual links to urban morphology are often shaky (Salvati et al., 2014; Zhang et al., 2023). Furthermore, most of the study focuses on large metropolitan regions, leaving out small and medium-sized cities, exurban and peri-urban areas where distinct urban form-sustainability dynamics may exist (Irwin & Bockstael, 2007; Zhang et al., 2023).

The review illustrates the ongoing debate over the best strategies for capturing and assessing urban morphology. Some specialists support complex spatial statistics, simulation models, and three-dimensional urban analytics, while others continue to use standard, accessible two-dimensional metrics (Batty & Longley, 1994; Chatzipoulka & Nikolopoulou, 2018; Zhang et al., 2023). The dichotomy between the desire for standardized, easily replicable measurements and the need for context-sensitive, nuanced analysis pervades much of the discussion in the field (Fleischmann et al., 2021; Zhang et al., 2023). Disputes about the implementation of fundamental concepts like compactness, diversity, and accessibility persist. These arguments highlight both discipline differences and the inherent complexity of urban systems (Batty, 2012).

This synthesis clearly demonstrates the ramifications for research, policy, and practice. The study topic would benefit from the development and widespread adoption of standardized taxonomies for urban morphology indicators and sustainability outcomes (Fleischmann et al., 2021; Zhang et al., 2023). Prioritizing spatial modeling tools, as well as high-resolution, three-dimensional spatial data, is critical for conducting a more rigorous analysis of the links between urban form and sustainability (Batty & Longley, 1994; Zhang et al., 2023). Interdisciplinary collaboration is required, particularly among urban morphology experts, landscape ecologists, and data scientists, to develop frameworks that integrate both the manmade and natural components of urban environments (Batty, 2012; Jabareen, 2006).

This analysis underlines the importance of urban planners and decision-makers systematically evaluating the metrics and data used in the development of sustainability solutions (Batty, 2012). Planners must recognize that commonly used indicators may not consistently represent local conditions or be directly relevant across settings, especially in the presence of data constraints or definitional differences (Fleischmann et al., 2021; Zhang et al., 2023). Policymakers in rapidly urbanizing areas must improve spatial data collecting and analysis capabilities to ensure that planning decisions are based on strong, context-relevant facts (Hu et al., 2021; Zhang et al., 2023).

This review contributes to the subject by providing a complete taxonomy of urban morphology measurements, highlighting existing limits and knowledge gaps, and proposing approaches for more rigorous and integrative research methodologies (Zhang et al., 2023). It encourages methodological and conceptual accuracy, increased geographic diversity in research, and a strong emphasis on the links between urban structure, natural qualities, and sustainable outcomes (Jabareen, 2006; Fleischmann et al., 2021; Salvati et al., 2014). The study consolidates existing data and identifies topics for more research, providing a foundation for future academic research and practical recommendations for practitioners and policymakers (Zhang et al., 2023).

As urbanization continues to affect landscapes and cultures, the importance of sustainable urban morphology research will only grow. Confronting existing restrictions and debates, while fostering greater consistency and integration, is critical to realizing urban morphology's potential as a tool for sustainable urban development (Jabareen, 2006; Zhang et al., 2023).

**Conclusion**

This comprehensive analysis consolidates the existing framework of spatial measurements and approaches in sustainable urban morphology, highlighting both significant advances and persistent issues (Zhang et al., 2023). A review of 89 peer-reviewed studies reveals that urban morphology is a complex and multifaceted discipline that encompasses over 365 distinct metrics across six primary dimensions: urban tissue configuration, street network, building-plot attributes, land use, natural features and greenspace, and urban expansion (Zhang et al., 2023). Despite this variation, the lack of standardization and uniformity in operational definitions is a significant hindrance to comparability across research and countries (Fleischmann et al., 2021; Jabareen, 2006).

Essential findings show that specific urban form patterns, such as compactness, high connectivity, and land use diversity, frequently support sustainability goals such as reduced land consumption, better mobility, improved energy efficiency, and increased social fairness (Jabareen, 2006; Song et al., 2017; Ye et al., 2015). The efficacy of specific metrics is frequently dependent on local context, scale, and their definition or implementation (Fleischmann et al., 2021; Nam et al., 2012). Research focuses mostly on cities in the Global North, leaving a huge gap in understanding the unique challenges and opportunities in rapidly urbanizing parts of the Global South (Hu et al., 2021; Zhang et al., 2023). Furthermore, natural traits and ecosystem services are underrepresented in urban morphology research, despite their growing importance in discussions about resilience, climate adaptation, and urban health (Salvati et al., 2014; Zhang et al., 2023).

Several practical recommendations emerge to advance the discipline. Researchers must emphasize the development and use of standardized taxonomies and operational definitions for urban morphology indicators (Fleischmann et al., 2021; Zhang et al., 2023). To better portray the complexities of urban environments, increased methodological rigor is required, notably using spatial modeling methodologies and three-dimensional data (Batty & Longley, 1994; Chatzipoulka & Nikolopoulou, 2018). Interdisciplinary collaboration in urban planning, landscape ecology, data science, and public health is critical for developing comprehensive frameworks that incorporate both built and natural features of urban environments (Jabareen, 2006; Batty, 2012).

The findings highlight the importance of practitioners thoroughly analyzing urban morphology measures used in planning and evaluation processes (Batty, 2012; Jabareen, 2006). Practitioners must be aware of the context-dependent nature of many metrics and seek to choose indicators that are both meaningful and actionable in their local area (Nam et al., 2012; Fleischmann et al., 2021). The use of extensive, spatially explicit data collection and analysis technology can improve planning outcomes and enable evidence-based decision-making (Batty & Longley, 1994; Zhang et al., 2023).

Policymakers are recommended to support steps that increase local capacity for spatial analysis, particularly in areas experiencing rapid urbanization (Hu et al., 2021; Zhang et al., 2023). Policies should promote data sharing and accessibility, foster interdisciplinary collaborations, and prioritize research on natural features and ecosystem services in urban environments (Salvati et al., 2014; Zhang et al., 2023). Cities may move closer to egalitarian, resilient, and environmentally sustainable growth by aligning urban policy with current facts on urban morphology and sustainability (Jabareen, 2006).

Future research must address the persistent underrepresentation of the Global South, widen comparative analyses, and study the links between urban form and sustainability across a wide range of socioeconomic and environmental contexts (Hu et al., 2021; Zhang et al., 2023). The discipline should advance by combining ecological health, landscape connectivity, and climate resilience measures into traditional urban morphology research (Salvati et al., 2014; Zhang et al., 2023). The development of high-resolution, three-dimensional datasets and the application of spatial modelling tools will considerably improve the trustworthiness of future research (Batty & Longley, 1994; Chatzipoulka & Nikolopoulou, 2018).

Significant questions remain about the best approaches for standardizing urban morphology metrics while allowing for contextual adaptation, more efficiently incorporating natural factors into urban form analysis, and translating research findings into scalable, practical policy (Fleischmann et al., 2021; Zhang et al., 2023). Addressing these questions will be critical for realizing the full potential of urban morphology research to inform sustainable urban transformation in the next decades (Jabareen, 2006; Zhang et al., 2023).